\documentclass[11pt]{article}
\usepackage[T1]{fontenc}
\usepackage[utf8]{inputenc}
\usepackage{lmodern}
\usepackage[letterpaper,margin=1in]{geometry}
\usepackage{microtype,amsmath,amssymb,graphicx,booktabs,tabularx,array}
\usepackage{xurl}
\usepackage[colorlinks=true,allcolors=blue]{hyperref}
\usepackage{caption}
\newcolumntype{Y}{>{\raggedright\arraybackslash}X}
\title{UAE GovTrust Scope Filtering Gains and Routing Failures in Bilingual Public Service Retrieval}
\author{Aleksandar Mihajlovic\\[3pt]\small Technical Faculty, Singidunum University, Belgrade, Serbia\\\small PhD student; based in Abu Dhabi, United Arab Emirates}
\date{21 October 2026}
\hypersetup{pdftitle={UAE GovTrust Scope Filtering Gains and Routing Failures in Bilingual Public Service Retrieval},pdfauthor={Aleksandar Mihajlovic}}
\begin{document}
\maketitle
\begin{abstract}
Public-service queries can mention several locations while only one controls the requested service. We study this distinction in UAE GovTrust, a controlled English-Arabic retrieval diagnostic comprising 22 authored service summaries and 50 scenarios with 110 query variants. Twelve methods separate lexical and dense ranking, rank fusion, scope filtering, a multilingual cross-encoder, and direct service lookup. On 72 matched answerable queries with bilingual evidence, role-filtered reciprocal rank fusion retrieves a relevant first result for 69 queries (95.8\%), versus 47 (65.3\%) for unfiltered fusion; filtered dense retrieval and direct lookup each achieve 68 (94.4\%). With English-only evidence, excluding zero-score lexical candidates raises filtered fusion from 52 to 64 correct first hits. However, the unchanged router selects the correct action and scope for only 2 of 168 post-audit perturbations covering paraphrases, irrelevant facts, negation, and corrections. These perturbations are deliberately authored diagnostics, not held-out evaluation. Fresh-environment execution reproduces all 1,320 original ranking orders and actions. The results show that correct scope metadata can improve ranking while brittle scope inference can remove the evidence needed to answer. The study evaluates retrieval and routing, not generated answers or legal correctness; independent Arabic and domain validation remains necessary.
\end{abstract}

\section{Introduction}
A public-service question may mention where a person lives, where a vehicle is registered, where a property is located, and which authority issued a document. These are different relations. Retrieving a passage about a city named in the question is therefore insufficient: the system must determine which relation controls the requested service. For example, in the catalogue studied here, a question about a toll gate in Abu Dhabi should retrieve the Abu Dhabi toll-system card even when the vehicle is registered in Dubai. Reversing those two locations changes the relevant card without changing the set of location names.

This problem matters to retrieval-augmented generation (RAG) because a generator cannot reliably ground an answer in evidence that an upstream filter has discarded. It is also relevant to enterprise service catalogues, policy portals, and operational assistants in which organizational units or service territories overlap. Those applications motivate the study; they are not deployment settings evaluated in the experiments. The present work isolates the retrieval and routing stages so their errors can be observed without a generator or an automated answer judge obscuring the source of failure.

UAE GovTrust is presented as a controlled diagnostic, not as an expert-validated benchmark of UAE law. Its corpus consists of short, authored service summaries, its questions are synthetic, and its labels describe relevance within that catalogue. An initial development experiment suggested a large benefit from scope filtering. A subsequent technical audit reproduced the original arithmetic but exposed failures in the rule-based router. This report incorporates that criticism through stronger comparisons, an explicit post-audit stress set, and a separation of routing errors from errors in ranking eligible evidence.

The contribution has three parts. First, it provides a reproducible bilingual diagnostic in which the service-controlling location differs from an incidental location. Second, it disentangles scope selection, ranking, sparse-branch eligibility, and service lookup through twelve comparison methods. Third, it documents how the benefit of a correct hard filter can coexist with severe failures of the mechanism that chooses that filter. The resulting evidence supports a limited engineering conclusion: scope inference must be evaluated independently of the retrieval gains obtained when the scope is correct. No claim is made to the first UAE regulatory RAG system, a new general filtering algorithm, or production readiness.

\section{Related research}
Regulatory retrieval and grounded legal question answering already have substantial prior work. RIRAG introduced ObliQA using Abu Dhabi Global Market regulatory material, and the RIRAG-2025 shared task developed that research setting \cite{r1,r2}. Version- and scope-aware question answering over normative documents has also been studied in a deployed system \cite{r3}. These studies rule out treating a jurisdiction or version filter itself as a new contribution. UAE GovTrust instead examines a small, explicitly controlled ambiguity: several locations are present, but they occupy different service-related roles. It does not reproduce the document scale, domain validation, or end-to-end evaluation of those systems.

Multilingual and bilingual evaluation are likewise established. bLLeQA examines grounded legal question answering in French and Dutch \cite{r4}; MIRAGE-Bench evaluates multilingual RAG systems \cite{r7}; and RAGBench distinguishes retrieval-related and answer-related quality dimensions \cite{r6}. Alrajeh studies Arabic-English citation fabrication, false premises, and unanswerable questions for Saudi data-protection law and the GDPR using manually verified references \cite{r15}. Wang and Suresh audit the institutional frameworks assumed by multilingual chatbots when jurisdiction is unspecified \cite{r14}. The present questions usually specify competing locations rather than omitting location altogether, and the outputs assessed here are ranked cards and gate decisions rather than legal citations or generated explanations.

The diagnostic design builds on behavioral testing. CheckList distinguishes targeted capabilities and invariance tests from aggregate accuracy \cite{r8}. FollowIR evaluates whether retrieval models follow relevance instructions rather than merely match their words \cite{r12}. NevIR tests ranking distinctions that depend on negation \cite{r13}. Accordingly, the role swaps, irrelevant-fact additions, and negation cases in this report are a domain-specific application of existing behavioral evaluation principles. They are not a new general testing paradigm. The Arabic text is intended as Modern Standard Arabic; it is not an Emirati-dialect benchmark, unlike the dialect coverage investigated by EDRAC \cite{r5}.

The retrieval comparisons use BM25 \cite{r10}, multilingual E5 embeddings \cite{r9}, reciprocal rank fusion (RRF) \cite{r11}, and a compact multilingual cross-encoder associated with mMARCO training \cite{r16,r17}. The chosen cross-encoder is a practical open-model comparison, not an assertion of the strongest available reranker. A direct service lookup is included because a 22-card catalogue may be solvable largely through service classification and scope selection. This control tests the need for retrieval complexity within the actual experimental setting.

\section{Materials and methods}
\subsection{Task and catalogue annotations}
The frozen catalogue contains 22 English-Arabic service cards linked to 21 distinct official-source URLs. Services cover business licensing and incorporation, tenancy and utilities, rental disputes, tolls and parking, buses, visa-validity channels, identity documents, VAT registration, work permits, and national digital identity. Applicability labels use five values: federal (FED), Abu Dhabi (AD), Dubai (DXB), Abu Dhabi Global Market (ADGM), and Dubai International Financial Centre (DIFC). These are catalogue-routing labels. They do not form a complete legal ontology. In particular, the publisher level and the service applicability label are separate fields: a federal publisher can describe a channel relevant to a particular issuing location.

The cards are AI-assisted factual summaries, not copies of official documents. Their source register records access status, observation date, publisher level, and unresolved temporal and redistribution questions. At construction, thirteen cards were based on inspected page bodies and nine on indexed excerpts. The revision inspected page bodies supporting three additional cards: Dubai tenancy registration, the Dubai Rental Disputes Center overview, and Federal Tax Authority VAT registration. Six cards remain supported only by indexed excerpts. The experimental text was kept frozen during this recheck. Inspecting a source page does not independently validate the Arabic wording, all legal exceptions, or the candidate relevance labels.

The original set has 50 scenarios: 36 answerable, eight requiring clarification, and six requiring abstention under the catalogue policy. Thirty answerable scenarios require one card and six require two. Every scenario has English and Arabic variants; ten also have a mixed-language variant. The primary retrieval analysis uses the 72 English-Arabic variants of the 36 answerable scenarios. The 100 matched variants of all scenarios assess gate decisions. The additional ten mixed-language variants are exploratory and are not pooled into the primary estimate. Fourteen paired families change the relevant local scope, while four families hold a federal service invariant under a location change.

Questions, relevance labels, and routing patterns were developed within the same workflow. They were not independently authored, and no untouched test split exists. The Arabic variants and domain labels remain candidates pending qualified independent review. All references to a correct scope or relevant card in the results therefore mean agreement with this explicit catalogue annotation, not an independent finding about UAE law or administrative eligibility. Scenarios that request historical rules, unavailable exact fees, individual dispute outcomes, or an unsupported territory deliberately exceed the frozen catalogue; their abstention labels do not imply that those questions are unanswerable by a larger system.

\begin{table}[!htbp]
\centering\small
\caption{Experimental units and evaluation sets. Translations and perturbations are dependent observations.}
\begin{tabularx}{\linewidth}{p{.24\linewidth}p{.10\linewidth}Y}
\toprule
\textbf{Component} & \textbf{Size} & \textbf{Use}\\
\midrule
Authored service cards & 22 & Bilingual or English-only evidence\\
Original scenarios & 50 & 36 answer; 8 clarify; 6 abstain\\
Original query variants & 110 & 50 English; 50 Arabic; 10 mixed\\
Primary retrieval set & 72 & Matched answerable variants\\
Gate-decision set & 100 & All matched English-Arabic variants\\
Post-audit controls & 24 & Six service families; two locations; two languages\\
Post-audit perturbations & 168 & Seven transformations of each control\\
\bottomrule
\end{tabularx}
\end{table}
\subsection{Scope filtering and ranking}
For query q, let G(q) be the set of annotated relevant cards. A router predicts an action in \{answer, clarify, abstain\} and an allowed-scope set. An answer action produces the candidate set C(q) of cards whose applicability labels are allowed; any other action produces an empty set. A ranker orders candidates and returns a list L(q). The current local-service policy retains FED alongside the selected local scope. This permits the separate federal VAT card in relevant multi-service questions; it is a deliberate catalogue policy rather than a universal rule of jurisdiction.

\begin{equation}
\operatorname{Recall}@k(q)\leq\frac{|G(q)\cap C(q)|}{|G(q)|}.
\end{equation}
The inequality follows directly because the retrieved top-k list is a subset of C(q). It expresses a basic constraint rather than a new retrieval theorem: if the gate suppresses retrieval or the filter excludes all relevant cards, no subsequent reranker can recover them. Consequently, action correctness, scope agreement, evidence survival, and conditional ranking quality are reported separately.

BM25 uses k1 = 1.2 and b = 0.75 with query term counts. Preprocessing lowercases text, removes Arabic diacritics and tatweel, and normalizes selected alif and ya forms. E5 uses intfloat/multilingual-e5-small with query and passage prefixes, attention-masked mean pooling, L2 normalization, and exact dot-product ranking. Inputs are truncated to 512 tokens; vectors have 384 dimensions. RRF sums 1/(60 + rank) over lexical and dense branches, with branch depth 100. All 22 cards fit within that depth. Ties are broken by the existing ascending document-ID order.

The original RRF implementation admits zero-score BM25 cards, so an uninformative lexical branch can still contribute ranks determined partly by document IDs. PositiveRRF removes cards with BM25 scores of zero from that branch before assigning ranks. The dense branch remains unchanged. Scope filtering precedes branch ranking in routed variants. This detail is particularly relevant when Arabic questions are matched to English-only evidence, where sparse lexical overlap can be limited.

MentionRRF filters by every recognized location mentioned in the query, retaining FED. RoleRRF instead calls the original deterministic infer function to select a service-controlling role and scope before RRF. OracleRRF substitutes the annotated action and allowed scopes; it is an upper-bound control for scope selection with that ranker, not a deployable method. RoleDense and RolePositiveRRF use the same unchanged router with dense ranking or positive-score fusion. The router was not patched after the audit or tuned to the new stress outcomes. Its exact ordering of global checks, regular expressions, and clause matching is released in the code.

CrossEncoder uses cross-encoder/mmarco-mMiniLMv2-L12-H384-v1. It jointly encodes each query-card pair and ranks the scalar logits without an activation. All 22 cards are scored, avoiding a first-stage recall bottleneck in this small corpus. RoleCrossEncoder uses the same logits but restricts the candidate set to the original router output. Accuracy calculation shares full-corpus pair logits across these two variants; timed routed requests score only eligible cards. Pair batches contain at most eight examples, with longest-first truncation at 512 tokens. Neither neural model is fine-tuned on UAE GovTrust.

DirectLookup combines the unchanged router with bilingual keyword patterns for the thirteen service types, then returns cards matching both the selected scope and detected service, in document-ID order. It does not encode or search card text. Its evidence-language invariance is therefore a design property, not evidence of cross-lingual document retrieval. The service patterns are a transparent catalogue-specific comparison and share the development bias of the routing rules. They can fail to identify a service even when the scope is correct.

\subsection{Behavioral diagnostics after the audit}
The nine previously disclosed audit probes were first reproduced as development examples: two retained the expected route and seven failed. They were not treated as unseen evaluation. A separate, frozen post-audit diagnostic set then covered six services: business licensing, tenancy registration, tolls, parking, buses, and visa validity. Each service has two location assignments and two languages, giving 24 unchanged controls. Seven transformations per control produce 168 perturbations: paraphrasing the controlling clause; adding an irrelevant Sharjah reference, a 2021 phone-purchase date, or an unknown old phone number; placing the distracting location after the controlling clause; negating an alternative location; and correcting an earlier location.

Labels are inherited from the target service and location in the original catalogue. Negation and correction cases explicitly identify the final intended location, but their expected labels have not received independent linguistic or domain review. The transformations were authored after the audit with knowledge of the router. They deliberately exercise known weaknesses, were frozen before the new comparative run, and are neither representative user traffic nor a held-out robustness benchmark. Their purpose is to expose mechanisms and compare behavior on disclosed cases, not estimate a deployment failure rate.

\subsection{Measures and statistical interpretation}
Hit@1 is one when the first retrieved card belongs to G(q), including either relevant card in a two-card question. Recall@5 is the fraction of G(q) in the first five results. nDCG@10 uses binary relevance with logarithmic discount and normalization by the ideal ordering. Empty rankings score zero on these retrieval measures. Wrong-scope first hits count returned first cards whose scope lies outside the annotated allowed set; an empty ranking is not counted as a wrong-scope return and is reported separately. This distinction prevents a system that refuses every query from appearing reliable merely because it returns no inapplicable cards.

Routing agreement requires the correct action and, for answerable cases, an exact match to the annotated allowed-scope set. Filter survival measures whether all relevant cards remain eligible. Error decomposition distinguishes incorrect gate actions, answering with an incorrect scope, and a wrong first result despite correct routing. A paired local-target measure additionally requires both members of a scope-changing family to retrieve the relevant local card first; a shared federal card does not satisfy that criterion. Translations and perturbations are not treated as independent samples.

The original RoleRRF-minus-RRF Hit@1 difference is accompanied by the preserved exploratory percentile bootstrap over eighteen paired scenario families: 10,000 resamples with seed 20261021. Its interval describes variation over the authored families and is not a population confidence statement. The new comparisons and stress outcomes are descriptive; no confirmatory significance claim, multiplicity-adjusted hypothesis test, or population generalization is made. Freezing a post-audit set improves traceability but does not make it independently held out.

\subsection{Execution and latency protocol}
The revision used a newly installed Python environment on the same host as the original study. Execution used an AMD EPYC 9V74 80-Core Processor CPU, one inference thread, float32 weights, and no quantization. Both models were downloaded from pinned revisions with recorded SHA-256 hashes. Fresh E5 embeddings were generated for both evidence-language conditions. Index construction and model loading were separated from request measurements. The model and package manifests, environment record, and complete dependency lock are included with the reproducibility material.

Latency was measured on the 72 original matched answerable queries with bilingual cards. Each method received three warm-up queries followed by three seeded, randomized passes, yielding 216 measurements per method and 2,592 in total. Timed requests included routing where used, tokenization, fresh query encoding or cross-encoder scoring, sorting, and output construction. There was no query-vector cache, network request, generator, concurrency test, or cold-start measurement. The median and empirical 95th percentile describe this single-host, 22-card workload; they are not enterprise service-level guarantees.

\section{Results}
\subsection{Original matched retrieval comparison}
Table 2 reports the 72 matched answerable queries with bilingual cards. RoleRRF retrieves a relevant first result for 69 queries, compared with 47 for RRF: a difference of 22 queries, or 30.6 percentage points. The exploratory family-bootstrap interval for this difference is 19.4 to 41.7 percentage points. All 47 RRF successes are also RoleRRF successes; RoleRRF adds 22 successes and shares three failures. Both methods can often retrieve relevant material within the first five results, so the first-rank difference should not be interpreted as an equal difference in complete evidence coverage.

\begin{table}[!htbp]
\centering\small
\caption{Retrieval on 72 matched answerable queries with bilingual cards. Wrong scope is the count of inapplicable first results; Hit@1 is a percentage. OracleRRF uses annotations.}
\begin{tabularx}{\linewidth}{Yrrrr}
\toprule
\textbf{Method} & \textbf{Hit@1 (\%)} & \textbf{Recall@5} & \textbf{nDCG@10} & \textbf{Wrong scope}\\
\midrule
BM25 & 61.1 & 0.951 & 0.814 & 27\\
E5 & 66.7 & 0.972 & 0.859 & 23\\
RRF & 65.3 & 0.951 & 0.853 & 25\\
MentionRRF & 65.3 & 0.958 & 0.854 & 25\\
RoleRRF & 95.8 & 0.986 & 0.972 & 0\\
OracleRRF & 95.8 & 0.986 & 0.972 & 0\\
PositiveRRF & 65.3 & 0.951 & 0.853 & 25\\
RoleDense & 94.4 & 1.000 & 0.974 & 0\\
RolePositiveRRF & 95.8 & 0.986 & 0.972 & 0\\
CrossEncoder & 43.1 & 0.889 & 0.738 & 39\\
RoleCrossEncoder & 86.1 & 0.986 & 0.936 & 0\\
DirectLookup & 94.4 & 0.944 & 0.942 & 0\\
\bottomrule
\end{tabularx}
\end{table}
The new controls qualify the apparent advantage. RoleDense and DirectLookup each achieve 68 correct first hits out of 72, close to RoleRRF. The tested cross-encoder achieves 31/72 without filtering and 62/72 with filtering. Thus this model does not establish an advantage over E5 on the diagnostic. The result concerns this exact model, corpus, and input format; it is not evidence that cross-encoders generally underperform dense retrieval.

The unchanged router matches all 100 original matched gate labels, and its scopes match the annotations on all 72 answerable variants. That perfect development-set agreement explains why RoleRRF and OracleRRF coincide here. It does not establish robust scope inference. The three RoleRRF first-rank failures are Arabic variants Q01, Q07, and Q34, where the selected scope is correct but the leading card describes the wrong service. DirectLookup has four empty-ranking failures, the English and Arabic variants of Q33 and Q35: its service keyword detector does not recognize those incorporation formulations. These failures occur after correct scope selection.

On the 28 language-specific pairs from fourteen scope-changing families, requiring both relevant local first results is stricter than ordinary Hit@1. RRF succeeds on 3/28 pairs and RoleRRF on 23/28. Mere mention filtering does not resolve which of two named locations controls the service. More generally, under the implemented bag-of-words BM25, queries with identical normalized term-count vectors have identical scores for every card. Location-role swaps that preserve those counts therefore cannot both be solved by changing BM25 ranking alone when their target local cards differ.

\subsection{Evidence language and sparse branch behavior}
Table 3 separates query language from evidence language. With bilingual cards, RoleRRF achieves 36/36 English and 33/36 Arabic first hits. With English-only cards, its Arabic result falls to 16/36, even though its scope predictions remain correct. This is a within-scope ranking problem rather than a failure to identify the location. Unfiltered E5 reaches 17/36 Arabic first hits in that condition, while ordinary RRF reaches only 7/36.

\begin{table}[!htbp]
\centering\small
\caption{Hit@1 percentages by query and evidence language. Each column contains 36 answerable queries. DirectLookup does not inspect evidence text.}
\begin{tabularx}{\linewidth}{Yrrrr}
\toprule
\textbf{Method} & \textbf{\shortstack{Bilingual /\\EN}} & \textbf{\shortstack{Bilingual /\\AR}} & \textbf{\shortstack{English-only /\\EN}} & \textbf{\shortstack{English-only /\\AR}}\\
\midrule
BM25 & 61.1 & 61.1 & 61.1 & 8.3\\
E5 & 75.0 & 58.3 & 72.2 & 47.2\\
RRF & 66.7 & 63.9 & 66.7 & 19.4\\
MentionRRF & 66.7 & 63.9 & 66.7 & 25.0\\
RoleRRF & 100.0 & 91.7 & 100.0 & 44.4\\
OracleRRF & 100.0 & 91.7 & 100.0 & 44.4\\
PositiveRRF & 66.7 & 63.9 & 66.7 & 47.2\\
RoleDense & 100.0 & 88.9 & 100.0 & 77.8\\
RolePositiveRRF & 100.0 & 91.7 & 100.0 & 77.8\\
CrossEncoder & 41.7 & 44.4 & 41.7 & 36.1\\
RoleCrossEncoder & 94.4 & 77.8 & 91.7 & 72.2\\
DirectLookup & 94.4 & 94.4 & 94.4 & 94.4\\
\bottomrule
\end{tabularx}
\end{table}
Excluding zero-score lexical candidates increases unfiltered fusion from 31/72 to 41/72 first hits with English-only cards. More substantially, RolePositiveRRF rises from RoleRRF's 52/72 to 64/72, matching RoleDense. The Arabic component rises from 16/36 to 28/36. This ablation identifies a concrete implementation hazard: ranking zero-overlap sparse candidates can introduce document-ID ordering into a fusion calculation. Scope filtering alone does not repair that ranking behavior. In the bilingual condition, PositiveRRF and RolePositiveRRF leave their respective ordinary-fusion first-hit counts unchanged.

\subsection{Routing under disclosed perturbations}
The stress evaluation changes the interpretation of the development results. The original router selects the correct action and exact scope on 2/168 perturbations (1.2\%). It makes 120 incorrect gate decisions and answers with an incorrect scope in 46 cases. All required evidence survives its candidate filter in only 2/168 cases. Of the two correctly routed cases, one still yields the wrong first result under RoleRRF. Both correctly routed perturbations are Arabic business paraphrases that retain a matching business-location pattern. Figure 2 separates these mechanisms rather than counting only wrong-scope returned cards.

RoleRRF retrieves a relevant first result for 1 of 168 perturbations, compared with 76/168 for RRF, 82/168 for E5, and 59/168 for the unfiltered cross-encoder. OracleRRF obtains 161/168 when supplied with the annotated action and scope. Figure 1 compares each method on the 24 unchanged controls and their 168 transformations; the complete counts appear in Appendix A. The original 72-query result and the stress result have different service mixtures, so their percentage difference is not treated as a matched population effect.

\begin{figure}[!htbp]
\centering
\includegraphics[width=0.94\linewidth]{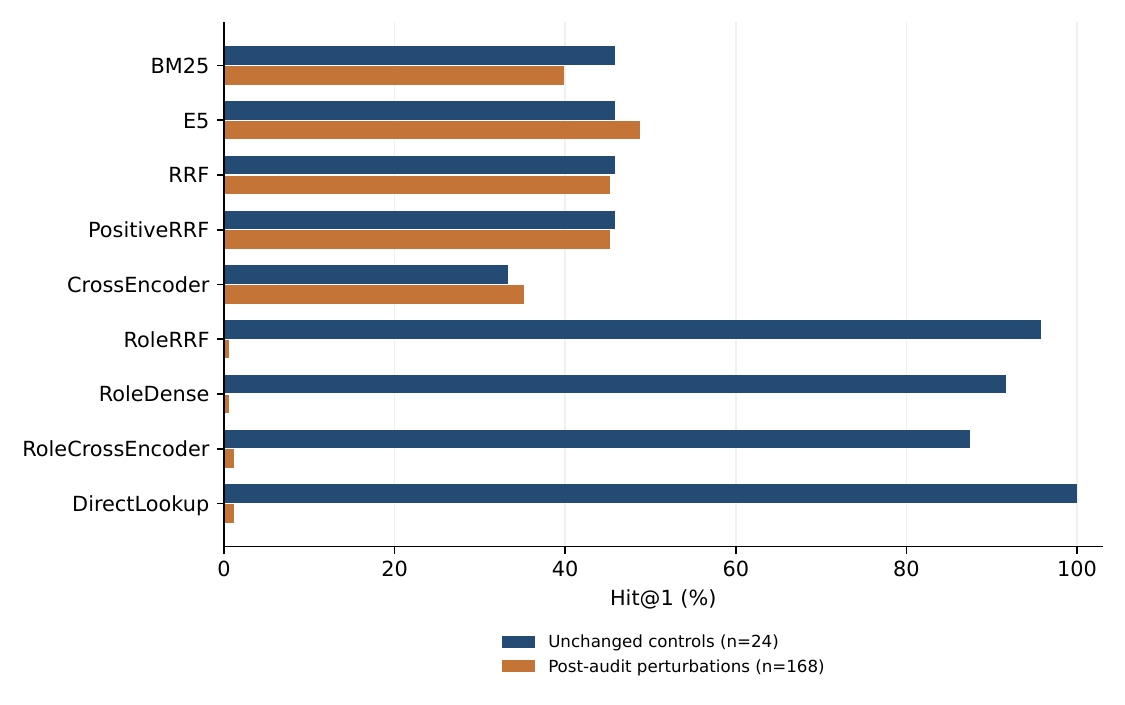}
\caption{First-result retrieval on 24 unchanged controls and seven perturbations of each control. The perturbations were deliberately authored after the audit; bars describe these examples, not general robustness estimates. Bilingual cards are used throughout.}
\end{figure}
The mechanisms are visible in the code. An occurrence of Sharjah triggers abstention before a role is resolved, even when it is an unrelated past residence. A year matching the historical rule can suppress an otherwise current service question. A generic missing-information phrase can trigger clarification about an irrelevant fact. The controlling-clause parser can include a trailing distracting location and declare a conflict. It does not systematically resolve negation or revise an earlier location after a correction. A downstream reranker cannot repair a gate that returns no candidates or an incorrect filter that excludes the required local card.

\begin{figure}[!htbp]
\centering
\includegraphics[width=0.94\linewidth]{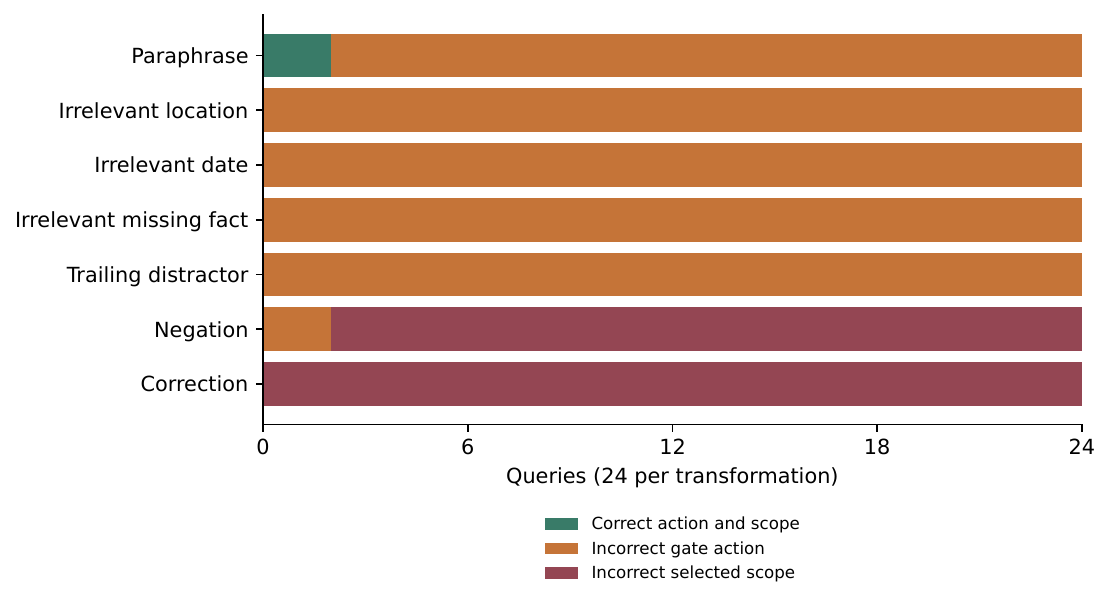}
\caption{Error decomposition for the unchanged router, shared by the non-oracle routed methods. Each transformation contains 24 queries. Correct routing requires both an answer action and the exact annotated allowed-scope set. The categories are mutually exclusive.}
\end{figure}
These failures do not imply that filtering is intrinsically harmful. OracleRRF shows the value of correct scope selection under the same perturbations, while the hard-filter recall bound explains why incorrect selection can dominate the end-to-end retrieval outcome. The stress set was designed with knowledge of these weaknesses, so its low score is evidence that the failures exist and recur across the included formulations. It is not an unbiased estimate of how often they occur in natural service enquiries.

\subsection{Latency and reproduction}
\begin{table}[!htbp]
\centering\small
\caption{Warm request latency on the 72 original answerable queries with bilingual cards: 216 measurements per method, one CPU thread. Units are milliseconds. Model loading, indexing and generation are excluded.}
\begin{tabularx}{\linewidth}{Yrr}
\toprule
\textbf{Method} & \textbf{Median (ms)} & \textbf{p95 (ms)}\\
\midrule
BM25 & 0.302 & 0.493\\
E5 & 36.003 & 48.321\\
RRF & 36.567 & 49.644\\
MentionRRF & 36.809 & 49.813\\
RoleRRF & 36.871 & 51.585\\
OracleRRF & 36.727 & 49.453\\
PositiveRRF & 37.646 & 50.754\\
RoleDense & 36.394 & 48.885\\
RolePositiveRRF & 37.324 & 49.419\\
CrossEncoder & 1916.310 & 2223.633\\
RoleCrossEncoder & 985.762 & 1199.124\\
DirectLookup & 0.222 & 0.428\\
\bottomrule
\end{tabularx}
\end{table}
In the revised run, the median request takes 36.57 ms for RRF, 36.87 ms for RoleRRF, and 0.222 ms for DirectLookup. The unfiltered cross-encoder takes 1916.31 ms, compared with 985.76 ms for its routed variant. The measurements reflect different work: direct lookup performs no neural inference, while cross-encoding processes query-card pairs. No claim is made about latency on large corpora, approximate nearest-neighbor indexes, concurrent traffic, or complete RAG requests.

Fresh-environment execution reproduces the ranking order and action of all 1,320 original records, with 0 mismatches. The maximum absolute change among the original saved-vector arrays is 0; the maximum recorded ranking-score change is 0. Independent metric arithmetic within the computational workflow checks all 4,944 revised predictions, and all 2,592 new timing records are retained. This is a clean-install reproduction on the same host, not an external replication or independent human validation.

\section{Discussion and limitations}
The central distinction is between having correct scope metadata and correctly inferring that metadata from a question. The original development templates make the second problem look solved, allowing a hard filter to expose its ranking benefit. The post-audit examples show that the assumption fails when irrelevant dates and locations, altered syntax, negation, or corrections enter the input. Reporting only oracle-filter performance, only development-template accuracy, or only wrong-scope returned results would conceal part of that failure. A useful evaluation must count non-answer errors and evidence exclusion alongside ranking errors.

The direct-lookup comparison also changes the architectural interpretation. A small catalogue with one or two cards per service-scope combination can often be handled by identifying a service and selecting its scope. Its strong score does not establish an alternative production solution, because its keyword detector shares the same development constraints and its router shares the stress failures. It does show that the catalogue does not justify a claim that a complex hybrid architecture is necessary. Retrieval research on full policies would need substantially more documents, overlapping services, chunk boundaries, duplicated and superseded provisions, and relevance judgments beyond catalogue membership.

The evidence-language experiment demonstrates a separate failure that persists even when routing is correct. A multilingual dense model can retrieve across languages, but a hybrid implementation can dilute that ability if its sparse branch promotes zero-overlap documents. The positive-score ablation is deliberately simple and reproducible. It should not be generalized into a claim that a particular fusion constant, weighting scheme, or sparse cutoff is optimal; none was tuned or compared systematically. Similar care is needed when interpreting the cross-encoder result: a single compact checkpoint and this synthetic catalogue do not represent the full range of reranking methods.

Several validity limits are fundamental. The corpus, translations, prompts, parser, and labels were co-developed with AI assistance. Independent Arabic and public-service experts have not validated them. The stress set is post hoc and intentionally adversarial to known rules. Sample counts reflect many dependent translations and transformations, with only six services in the stress design. Neither a large number of perturbations nor a narrow computational confidence interval compensates for that limited variety. Model pretraining exposure to public service descriptions is unknown. No independent test set, human relevance annotation study, or inter-annotator agreement estimate is available.

The material is also a service catalogue rather than a collection of official source passages. Six cards still lack an inspected full page body, and observing a current webpage does not establish an authentic historical version or a legally complete description. No experiment here evaluates document-authority verification, temporal change detection, generated-answer faithfulness, citation correctness, or legal eligibility. The hypothetical historical queries only test refusal when the frozen catalogue lacks a requested version. They cannot support claims about handling real policy changes. The report also does not establish safety, fairness, or production effectiveness for a government service.

A next-stage study should separate construction from evaluation. Qualified Arabic and domain reviewers should first check the source-backed labels and document exceptions. A different authoring group should then create unseen questions, preserving families across splits and withholding negation, correction, and irrelevant-fact formulations from parser development. Learned or schema-constrained language-model role extraction, query rewriting, and clarification strategies should be compared against the disclosed rules. Only after retrieval and routing are validated should a fixed generator be added and scored for supported claims, completeness, and citation applicability. These are proposed extensions; no such experiments are claimed in this report.

\section{Conclusion}
UAE GovTrust shows both the benefit and the fragility of scope filtering in a controlled bilingual public-service catalogue. With the original templates, role-filtered fusion improves first-result retrieval, but a simple lookup and filtered dense retrieval achieve similar results. Removing zero-score lexical candidates materially improves the English-only evidence condition. Under disclosed post-audit perturbations, the unchanged router frequently blocks retrieval or selects an inapplicable scope, reversing the practical benefit of filtering. The supported contribution is a reproducible diagnostic and error decomposition, not an expert-validated UAE benchmark or an end-to-end RAG system. Future evidence should test scope inference on independently authored questions before attributing trustworthiness to a metadata filter.

\section*{Declarations}
\subsection*{Author affiliation and funding}
Aleksandar Mihajlovic is the sole author. The work is associated with his PhD studies at the Technical Faculty, Singidunum University, Belgrade, Serbia. He lives and works in Abu Dhabi, United Arab Emirates. The study was self-funded from the author's own funds. No external funding, research grant, or commercial sponsorship supported the study.

\subsection*{Contributions and use of AI}
Aleksandar Mihajlovic identified the industry motivation and research questions, directed the investigation, reviewed the research materials, and assumes responsibility for the study design, interpretation, and manuscript. OpenAI ChatGPT/Codex was used extensively under his direction for literature discovery, drafting the English-Arabic diagnostic material, writing and executing experimental code, computational analysis and checks, and manuscript preparation. AI assistance is not represented as independent Arabic or domain validation. No AI system is listed as an author, and no generated answers or LLM-as-judge scores are experimental outcomes. The human author remains responsible for the accuracy and integrity of the submitted work.

\subsection*{Employment and competing interests}
The author is employed as Associate Director - Forward Deployed Engineer at Modon Holding, Abu Dhabi, United Arab Emirates. This employment is disclosed for transparency. The experiments use public-source-derived authored summaries and synthetic questions and contain no employer data. The research findings do not represent an official position or endorsement of Modon Holding, Singidunum University, or any UAE public authority.

\subsection*{Ethics and data availability}
No human participants were recruited, no private administrative records were used, and no personal service cases were evaluated. Human-participant consent is therefore not applicable to these computational experiments; no formal institutional exemption is claimed. The accompanying ancillary material contains the frozen diagnostic inputs, source registers, raw rankings, timings, analysis scripts, manifests, and reproduction instructions. Original source code is provided under the MIT license. Government webpage bodies and external model weights are not redistributed; their rights and licenses remain with the respective owners. Model identifiers, pinned revisions and checksums are supplied for separate retrieval.

\section*{Appendix A Complete stress set results}
The controls repeat 24 original queries; the perturbations contain 168 variants of those same service-location-language combinations. All use bilingual cards. Empty is the number of perturbations with no returned card. Wrong scope counts only returned inapplicable first cards. OracleRRF is an annotation-assisted reference. The full per-query action, scope, reason, ranking, and metric records accompany the paper.

\begin{table}[!htbp]
\centering\small
\caption{Complete stress-set first-result counts. The denominators are 24 controls and 168 perturbations.}
\begin{tabularx}{\linewidth}{Yrrrr}
\toprule
\textbf{Method} & \textbf{Control hits} & \textbf{Perturbed hits} & \textbf{Wrong scope} & \textbf{Empty}\\
\midrule
BM25 & 11/24 & 67/168 & 96 & 0\\
E5 & 11/24 & 82/168 & 86 & 0\\
RRF & 11/24 & 76/168 & 92 & 0\\
MentionRRF & 11/24 & 76/168 & 92 & 0\\
RoleRRF & 23/24 & 1/168 & 46 & 120\\
OracleRRF & 23/24 & 161/168 & 0 & 0\\
PositiveRRF & 11/24 & 76/168 & 92 & 0\\
RoleDense & 22/24 & 1/168 & 46 & 120\\
RolePositiveRRF & 23/24 & 1/168 & 46 & 120\\
CrossEncoder & 8/24 & 59/168 & 88 & 0\\
RoleCrossEncoder & 21/24 & 2/168 & 46 & 120\\
DirectLookup & 24/24 & 2/168 & 46 & 120\\
\bottomrule
\end{tabularx}
\end{table}
\section*{Appendix B Reproduction and provenance}
The ancillary directory contains a reproducibility project with the original frozen data and version 0.1.1 outputs, alongside revision/data and revision/results for version 0.2.0. The unchanged original implementation is src/benchmark.py. The revision entry points are \path{revision/src/benchmark_revision.py}, \path{revision/src/analyze_revision.py} and \path{revision/src/reproduce_audit_probes.py}. The stress builder, pre-execution revision plan, dependency lock, model manifests, source recheck, and SHA-256 inventory preserve the construction and execution record. The stress data already supplied should be used for reproduction; rebuilding them creates a new freeze timestamp even when the text is unchanged.

E5 is pinned to revision 614241f622f53c4eeff9890bdc4f31cfecc418b3. The cross-encoder is pinned to revision 1427fd652930e4ba29e8149678df786c240d8825. Both are run in evaluation mode with float32 weights. The full environment file records exact package versions, token limits, CPU, thread setting and index construction times. Reproduction requires downloading the model files separately and matching their hashes. Timing values will depend on the host and load even when ranking results agree. The technical audit is represented by reproducible probe inputs and outputs; its private report is not part of the arXiv ancillary material.

\end{document}